\documentclass[conference, a4paper]{IEEEtran}
\IEEEoverridecommandlockouts
\usepackage{cite}
\usepackage{amsmath,amssymb,amsfonts}
\usepackage{algorithmic}
\usepackage{graphicx}
\usepackage{textcomp}
\usepackage{xcolor}
\usepackage{svg}
\usepackage{forest}
\usepackage{adjustbox}
\usepackage{comment}
\usepackage{tcolorbox}
\usepackage{tikz}
\usepackage{amssymb}
\usepackage{graphicx}
\usepackage{booktabs}
\usepackage{multirow}
\usepackage{tabularx}
\usepackage{makecell}
\usepackage[caption=false]{subfig}
\usepackage{booktabs} % \toprule, \midrule, \bottomrule, \cmidrule
\usepackage{siunitx}  % \SI and \si
\usepackage{url}
\usepackage{tikz}
\usepackage{eso-pic}

\usetikzlibrary{
    calc,
    positioning,
    arrows.meta,
    decorations.pathreplacing
}

\usetikzlibrary{
    positioning,
    calc,
    arrows.meta,
    decorations.pathreplacing
}

\def\BibTeX{{\rm B\kern-.05em{\sc i\kern-.025em b}\kern-.08em
    T\kern-.1667em\lower.7ex\hbox{E}\kern-.125emX}}
\begin{document}

\title{Efficient HQC Implementations on RISC-V: \\ A Cross-Layer Design Space Exploration
\thanks{This work was partly funded by the German Federal Ministry of Research,
Technology and Space as part of the ``SUSTAINET-GuarDian'' project under
grant number 16KIS2248, and co-funded by the European Union and the State of
Rhineland-Palatinate as part of the ``InnoProm'' project. It was also
co-funded by the Deutsche Forschungsgemeinschaft (DFG, German Research
Foundation) as part of ``GRK 2982''.}
}

\author{
\IEEEauthorblockN{Maximilian Schöffel, Johannes Feldmann, Elias Biehl and Norbert Wehn}
\IEEEauthorblockA{\textit{Microelectronic Systems Design Research Group} \\
\textit{RPTU Kaiserslautern-Landau}\\
Kaiserslautern, Germany \\
\{m.schoeffel,j.feldmann,biehl,norbert.wehn\}@rptu.de} 
}

\maketitle

\begin{abstract}
RISC-V-based hardware/software co-design has become an established methodology for implementing post-quantum cryptography (PQC) on embedded devices. Its software programmability, extensible instruction-set architecture, and optional dedicated accelerators create a broad design space that offers the flexibility required for crypto-agility and post-deployment updates.
For HQC, a PQC key-encapsulation mechanism selected for standardization, several RISC-V-based implementations have already been published. 
However, these works explore only a small subset of this design space.

We present, to the best of our knowledge, the most comprehensive exploration of RISC-V-based HQC. Our exploration is centered around a new, parametrizable architecture framework on which we characterize 29,920 distinct architectural configurations across all three HQC security levels on FPGA.
The Pareto-optimal configurations from this framework outperform the state-of-the-art RISC-V-based HQC implementation, ranging from 4.8$\times$ lower LUT utilization at 17\,\% lower latency to 69.6$\times$ lower latency at 30\,\% lower LUT utilization, while also requiring up to 10.6$\times$ fewer block RAMs.
\end{abstract}


\AddToShipoutPictureFG*{%
  \begin{tikzpicture}[remember picture,overlay]
    \node[
      anchor=south,
      align=center,
      text width=0.90\paperwidth,
      font=\footnotesize
    ] at ([yshift=5mm]current page.south) {%
      This paper has been accepted for publication at the 2026 IEEE NorCAS conference. \\
      \copyright~2026 IEEE. Personal use of this material is permitted.
      Permission from IEEE must be obtained for all other uses, in any current
      or future media, including reprinting/republishing this material for
      advertising or promotional purposes, creating new collective works,
      for resale or redistribution to servers or lists, or reuse of any
      copyrighted component of this work in other works.
    };
  \end{tikzpicture}%
}

\section{Introduction}
Since quantum computers are expected to break current public-key cryptosystems within the next decade~\cite{piani2024quantum}, the U.S. National Institute of Standards and Technology (NIST) has meanwhile selected the first Post‑Quantum Cryptography (PQC) standards.
Consequently, research attention is increasingly shifting toward the practical and efficient implementation of these algorithms in real-world systems.

So far, research on PQC implementations for embedded devices has predominantly focused on lattice-based schemes due to their relatively small keys, lower computational complexity, and early selection for standardization by NIST~\cite{nist8413upd1}. Following NIST's selection of the code-based Hamming Quasi-Cyclic (HQC) key-encapsulation mechanism (KEM) as an additional standard in 2025~\cite{NIST.IR.8545}, however, there has been a growing interest in efficient HQC implementations. Compared with lattice-based alternatives, HQC imposes substantially higher computational demands, making its implementation on resource-constrained embedded platforms particularly challenging.

However, beyond meeting constraints such as latency (the time to complete a cryptographic operation) and Field Programmable Gate Array (FPGA) resources, PQC implementations must also provide sufficient flexibility to support crypto-agility, i.e., the post-deployment adaptability of cryptographic implementations if vulnerabilities are discovered. Therefore, hardware/software co-design based on the open-source Instruction Set Architecture (ISA) RISC-V~\cite{riscv-isa} has become a widely used strategy for PQC, combining software programmability, a wide range of standard and custom ISA extensions, and hardware acceleration for computational bottlenecks.

For example, Fritzmann et al.~\cite{fritzmann2020risq} introduced multiple custom instructions targeting lattice-based schemes from the NIST process. In~\cite{schoeffel2022hqc}, the first hardware/software co-design for HQC followed a three-stage approach comprising software optimization, custom instruction-set extensions, and loosely coupled accelerators for FPGA and application-specific integrated circuit (ASIC) implementations. More recently, Dolmeta et al.~\cite{dolmetaTYRCARISCVTightlyCoupled2025} and Seo and Kim~\cite{seoHardwareSoftwareCoOptimization2026} targeted similar HQC computational kernels using custom RISC-V instructions. Additional work has investigated fully hardware-based HQC implementations~\cite{aguilar-melchorAutomatingCryptographicHardware2023,antognazzaHighEfficiencyHardware2024} and optimized implementations of individual computational kernels~\cite{schoeffel2023hqc,heLOCSLOwLatencyConStantTiming2023,tuLEAPLightweightEfficient2023}.

Despite these advances, prior work explores only a few manually selected configurations, potentially excluding architectures with better trade-offs, and does not systematically analyze the interactions among algorithmic refinements, RISC-V ISA choices, hardware/software partitioning, and accelerator design. Moreover, state-of-the-art (SoA) works primarily optimize latency or report only clock cycles~\cite{seoHardwareSoftwareCoOptimization2026}, providing limited insight into latency--resource trade-offs.

Our new contributions are:
\begin{enumerate}
    \item We employ a cross-layer methodology spanning algorithmic and implementation choices to develop a highly configurable RISC-V-based HQC architecture template. It supports all HQC security levels and allows variation of processor width and base ISA, standard and HQC-specific ISA extensions, interconnect width, HW/SW kernel mapping, accelerator configuration, and parallelism. In total, we cover 29,920 distinct architectural configurations.
    
    \item We characterize the FPGA resources and latency of the KEM operations for all these configurations on Xilinx Artix 7.
    We present a Pareto-based evaluation methodology that (i) identifies the lookup table (LUT)--latency optimal configurations and (ii) introduces metrics to determine how individual architectural choices shape and persist along the Pareto front.

    \item We show that the proposed architecture template enables a fine-grained architecture selection under application-specific FPGA resource and latency constraints. Compared with SoA RISC-V-based HQC implementations, every configuration on our Pareto front requires fewer FPGA resources and achieves lower latency than the most recent RISC-V-based HQC design.
    
\end{enumerate}
The remainder of this paper is structured as follows.
Section II briefly describes the background of HQC and KEMs.
Section III presents an analysis of the cross-layer design space, divided into the algorithmic and implementation layers.
Section IV introduces an evaluation methodology for analyzing the configurations.
Section V presents our results, while Section VI draws the conclusions.

\section{Background}
KEMs are a type of public-key cryptography that uses \textsc{KeyGen}, \textsc{Encaps}, and \textsc{Decaps} to establish a shared secret $ss$ between two entities that subsequently serves as keying material for symmetric cryptography like Advanced Encryption Standard (AES).
The operating principle for two entities, Alice and Bob, is as follows:
\begin{enumerate}
    \item \textbf{Key Generation}: Alice randomly generates the keypair ($s_{k_{\text{Alice}}}, p_{k_{\text{Alice}}}$) using \textsc{KeyGen}.
    \item \textbf{Encapsulation}: Bob uses $p_{k_{\text{Alice}}}$ and \textsc{Encaps} to generate the ciphertext $c$ and the shared secret $ss$.
    \item \textbf{Decapsulation}: Alice uses $s_{k_{\text{Alice}}}$ and \textsc{Decaps} to derive the same shared secret $ss$ from $c$.\\
\end{enumerate}
The code-based HQC KEM consists of four \textbf{kernels}~\cite{hqc5}:
\begin{enumerate}
    \item \textbf{Polynomial Arithmetic.} Arithmetic operations (addition, multiplication, and modular reduction) are performed over the quotient ring  $\mathcal{R} = \mathbb{F}_2[x]/(x^n-1)$, where $n$ depends on the code length. Polynomials can be represented in two ways: an \textbf{explicit} bit array  $v_i$ in $v=\sum_{i=0}^{n-1} v_i \cdot x^i$ where bit $i$ represents coefficient $v_i$, or a \textbf{support form} array $v=\sum_{i=0}^{\omega-1} x^{c_i}$ that holds the positions of the $\omega$ non-zero coefficients.
    \item \textbf{FIPS~202.} This standard encompasses the Secure Hash Algorithm 3 (SHA-3) functions and the SHAKE extendable-output functions~\cite{nationalinstituteofstandardsandtechnologyusSHA3StandardPermutationbased2015a}.
    Both are based on the \textsc{Keccak} sponge construction, which absorbs inputs into an internal state, applies the \textsc{Keccak}-f1600 permutation, and squeezes outputs from this state. HQC uses this for hashing and pseudorandom-data generation.
    \item \textbf{Error-Correcting Codes.} HQC combines two code constructions. A decodable code $\mathcal{C}$ with error-correction capability $\delta$ is constructed from concatenated Reed--Muller (RM) and Reed--Solomon (RS) codes and is used to recover the encoded message. In addition, a random double-circulant code defined by the parity-check matrix $\mathbf{H}$ introduces structured noise that can be corrected by $\mathcal{C}$~\cite{aguilar2018efficient}.
   \item \textbf{Sampling.} HQC expands random seeds using SHAKE to generate error polynomials with a fixed Hamming weight $\omega$. These polynomials provide the noise required for security.
\end{enumerate}

In this work, we consider the HQC specification 5.0 and support all three security levels~\cite{hqc5}.

\section{Design Rationale}
We chose RISC-V as the base platform because it combines software programmability with an extensible ISA and dedicated hardware acceleration, allowing us to explore fine-grained trade-offs between latency and FPGA resources.
Building on these properties, our methodology follows a cross-layer approach that explores a multidimensional design space divided into two layers. At the \emph{algorithmic layer}, alternative realizations of the HQC kernels are explored with respect to computational complexity and memory footprint. At the \emph{implementation layer}, we investigate different hardware/software partitioning and architectures on the RISC-V platform.

\subsection{Algorithmic Layer} 
Our platform does not use caches or off-chip memory, ensuring constant-time memory access. This property allows us to prioritize computationally efficient support-form algorithms for the sparse polynomials with low Hamming weight that would otherwise leak secret timing side-channel information on cached processors.

\subsubsection{Polynomial Arithmetic}
Polynomial multiplication is the primary bottleneck of HQC~\cite{schoeffel2022hqc}. Three realizations have been proposed: (i) \emph{window-based}~\cite{schoeffel2022hqc}\cite{aguilar-melchorAutomatingCryptographicHardware2023} reducing complexity from  $\mathcal{O}(n^2)$ (naive approach) to $\mathcal{O}(\omega n)$, (ii) \emph{rotation-based}~\cite{tuLEAPLightweightEfficient2023}, also $\mathcal{O}(\omega n)$, and (iii) \emph{Karatsuba/Toom--Cook}~\cite{hqc5}, which offers platform-independent constant-time execution but higher complexity than sparse-dense methods. We use the window-based method, as it enables preloading sampled error polynomials into the result buffer to perform post-multiplication addition implicitly.
Furthermore, this reduces storage requirements from $\mathcal{O}(n)$ to $\mathcal{O}(\omega)$ for sparse polynomials.

\subsubsection{FIPS 202}
The bottleneck is the \textsc{Keccak} permutation. We explore a full-state approach (contiguous 1600-bit blocks) and bit-sliced variants~\cite{bertoni1001WaysImplement}, which trade off latency, parallelism, and FPGA resources. Additionally, we evaluate sponge interface granularity (byte vs. 64b block) where the software handles the alignment to determine the impact on buffering and hardware/software transfer overhead.

\subsubsection{Error-Correcting Codes (ECC)}
While the decoding procedures are the primary computational bottlenecks of the ECC kernel, they are critical to the scheme's correctness. Deviations from the specified decoding logic may increase the decryption-failure probability and invalidate the formal security assumptions of HQC. Consequently, we restrict our design space to the reference algorithms~\cite{hqc5}.

\subsubsection{Sampling}
HQC 5.0 employs rejection-based sampling for \textsc{KeyGen} and Fisher-Yates (FY) for \textsc{Encaps}/\textsc{Decaps}. 
To optimize the FY uniqueness check, we utilize explicit-form lookups rather than the reference support-form search, reducing search complexity from $\mathcal{O}(n)$ to $\mathcal{O}(1)$~\cite{schoeffel2023hqc}.

\subsection{Implementation Layer}
\begin{figure}[t]
    \centering
    \includegraphics[
        width=\columnwidth
    ]{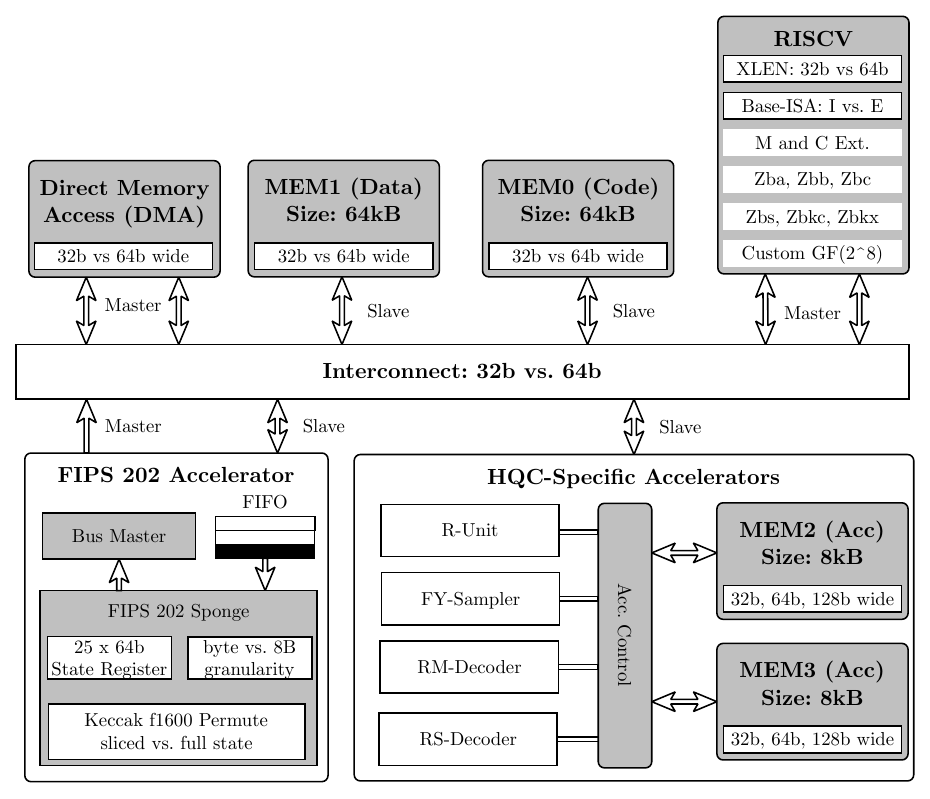}
    \caption{Configurable RISC-V architecture used for the exploration.
    The HQC-specific units share the dual-port memories MEM2 and MEM3, whereas \textsc{Fips}~202 is
    integrated as an independent bus master.}
    \label{fig:architecture}
\end{figure}

The architecture template for the explored implementations is shown in Fig.~\ref{fig:architecture}. It comprises the configurable RISC-V processor, code and data memories, an interconnect, an optional \textsc{Fips}~202 accelerator, and optional HQC-specific accelerators. Each HQC kernel can be either implemented in software or as a dedicated hardware accelerator; our unified architecture supports all three HQC parameter sets.

\subsubsection{Processor and ISA Configuration}
We use our proprietary RISC-V processor with a three-stage pipeline (fetch, decode, execute/memory), an instruction-prefetch depth of four, and support for four outstanding read requests.
Across the explored HQC software configurations, the processor mostly achieves between 1.35 and 1.60 cycles per instruction (CPI).

The processor configurations vary \texttt{XLEN}, the \texttt{RV32I}/\texttt{RV32E} base ISA, interconnect width, the \texttt{M} and \texttt{C} extensions, and four cumulative ISA profiles. The \texttt{base} profile comprises \texttt{Zicsr} and \texttt{Zifencei}; \texttt{zbasic} additionally includes \texttt{Zba}, \texttt{Zbb}, \texttt{Zbc}, and \texttt{Zbs}; \texttt{zbcrypto} further adds \texttt{Zbkc} and \texttt{Zbkx}; and \texttt{zpqc} includes the custom $GF(2^8)$ extension used by the RS codes~\cite{schoeffel2022hqc}. The \texttt{M} and \texttt{C} extensions are varied independently.
The floating-point \texttt{F} extension is excluded as HQC relies on integer arithmetic; the vector extension \texttt{V} is omitted due to its resource overhead, but remains a candidate for future work.

We prioritize these standardized extensions specified in~\cite{riscv-isa} over custom instructions because of their reuse potential across software workloads. The published custom extensions are omitted as they target algorithmic realizations outside of the selected kernels~\cite{dolmetaTYRCARISCVTightlyCoupled2025} or require the \texttt{F} extension~\cite{fritzmann2020risq}, providing limited benefit relative to dedicated acceleration. 

\subsubsection{Accelerator Organization}
Inter-kernel data dependencies within the KEM operations limit opportunities for concurrent execution.
Therefore, we closely follow our design described in~\cite{schoeffel2022hqc}: the \textsc{R-Unit} (polynomial arithmetic), FY sampler, and RM/RS decoders share two local memories. Only one compute unit is active at a time, allowing the memories and interconnect logic to be reused, and to keep intermediate values within the accelerator subsystem. The internal
parallelism $P\in\{32,64,128\}$ is coupled to the memory-width.

\textsc{Fips}~202 is integrated separately because of its streaming dataflow.
It operates as a bus master and writes pseudorandom data or hash outputs directly to the destination memory after being configured by the processor. Its architecture varies the sponge interface and the implementation of the \textsc{Keccak} permutation, comprising sliced variants based on~\cite{bertoniKeccak2015} and our full-state implementation.

\section{Evaluation Methodology}
Table~\ref{tab:configuration-space} summarizes the explored parameters yielding 29,920 functionally distinct architectures.
KEMs are typically employed with ephemeral keys, and \textsc{KeyGen} is also significantly less computationally intensive than \textsc{Encaps} and \textsc{Decaps} for HQC~\cite{schoeffel2022hqc}.
Therefore, each architecture is evaluated for the Alice metric (\textsc{KeyGen} + \textsc{Decaps}) and the Bob metric (\textsc{Encaps}, therefore without RM- and RS-Decoder as unused during \textsc{Encaps}), using the HQC-1, HQC-3, and HQC-5 parameter sets.
For the Alice metric, this yields $(64+16)\times34\times11 = 29{,}920$, and for the Bob metric this yields 6,160 configurations per security level, resulting in 198,000 to be evaluated architecture--operation combinations.

\begin{table}[t]
\centering
\caption{Explored architectural configuration parameters. $P$ refers to the internal parallelization of the HQC accelerators, and to the width of MEM2 and MEM3. Numbers in parentheses denote distinct configurations.}
\label{tab:configuration-space}
\resizebox{\columnwidth}{!}{%
\begin{tabular}{lll}
\toprule
\textbf{Group} & \textbf{Parameter} & \textbf{Alternatives} \\
\midrule

\multirow{5}{*}{\shortstack[l]{\textbf{RISC-V}\\$(64+16)$}}
 & XLEN
 & 32, 64 \\

 & Interconnect width
 & 32/64b (RV32), 64b (RV64) \\

 & Base ISA
 & I or E for RV32, I for RV64 \\

 & Optional extensions
 & none, M, C, M+C \\

 & ISA profile
 & \texttt{base}, \texttt{zbasic}, \texttt{zbcrypto},
   \texttt{zpqc} \\
\midrule

\multirow{4}{*}{\shortstack[l]{\textbf{HQC}\\$(34)$}}
 & R-Unit / FY
 & disabled, R-Unit, R-Unit + FY \\
 & Reed--Muller decoder
 & disabled, enabled \\

 & Reed--Solomon decoder
 & disabled, enabled \\

 & $P$
 & 32, 64, 128 bit \\
\midrule

\multirow{3}{*}{\shortstack[l]{\textbf{FIPS~202}\\$(11)$}}
 & Accelerator
 & disabled, enabled \\

 & Sponge interface
 & byte-aligned, block-aligned \\

 & \textsc{Keccak}-Permute
 &  4, 8, 16 or 32 slices, full state \\
 \bottomrule
\end{tabular}%
}
\end{table}

To analyze this large number of configurations on a Pareto front, let $a_j$ be any architectural alternative between the described configurations, e.g., the presence or absence of the \textsc{R-unit}.
Let $\mathcal{P}=(p_1,\ldots,p_N)$ be the set of Pareto-optimal configurations ordered from highest to lowest latency. For any $a_j$, let $x_i(a_j)\in\{0,1\}$ indicate whether $p_i$ selects $a_j$, and define
\begin{equation*}
    \mathcal{I}_j
    =
    \{\,i\in\{1,\ldots,N\}\mid x_i(a_j)=1\,\}.
\end{equation*}
For $\mathcal{I}_j \neq \varnothing$, define $i_j^{*}=\mathrm{min}(\mathcal{I}_j)$ as the index of the first Pareto point that uses $a_j$, and $i_j^{\dagger}=\mathrm{max}(\mathcal{I}_j)$ as the index of the last Pareto point that uses $a_j$.
We quantify the prevalence and continuity of $a_j$ through the occupancies
\begin{equation*}
    \eta_{\mathrm{tot}}(a_j)
        = \frac{|\mathcal{I}_j|}{N}, \qquad 
    \eta_{\mathrm{rel}}(a_j)
        = \frac{|\mathcal{I}_j|}
        {i_j^\dagger-i_j^*+1}.
\end{equation*}

The total occupancy $\eta_{\mathrm{tot}}(a_j)$ measures the prevalence of an architectural feature over the complete Pareto front; a value close to $1$ indicates that the feature is required in almost all Pareto-optimal configurations.
The relative occupancy $\eta_{\mathrm{rel}}(a_j)$ measures the continuity of a feature between its first and last occurrence on the Pareto front. A high $\eta_{\mathrm{rel}}(a_j)$ together with a low $\eta_{\mathrm{tot}}(a_j)$ therefore indicates that the feature is relevant only within a limited region of the front, but is selected consistently throughout that active region.

Let $\mathcal{R}_j$ form the maximal contiguous intervals contained in $\mathcal{I}_j$ with the number of intervals
\begin{equation*}
    r_j=|\mathcal{R}_j|.
\end{equation*}
A large $r_j$ therefore indicates frequent changes in the selection of the respective architectural feature along the Pareto front, i.e., the feature is repeatedly activated and deactivated and thereby contributes to local LUT--latency refinements.

The persistence range $\mathcal{Q}_j$ is the maximal occurrence run ending at the minimum-latency point $p_N$, with normalized length
\begin{equation*}
    \eta_{\mathrm{pers}}(a_j)
    =
    \frac{|\mathcal{Q}_j|}{N}.
\end{equation*}

If $\eta_{\mathrm{pers}}(a_j)>0$, the respective feature is required
at the minimum-latency Pareto-optimal point. A larger $\eta_{\mathrm{pers}}(a_j)$ indicates that the feature is required earlier along the Pareto front and remains selected up to the minimum-latency endpoint.

\section{Results}
The software part was implemented in \textsc{C} based on the HQC reference~\cite{hqc5}, optimized for the selected processor width and ISA extensions, and the software was compiled using GCC Version 15.2.1 with the \texttt{-Os} flag.
For the \textsc{Fips}~202 software we used the RISC-V optimized implementations from the XKCP project for \texttt{RV32} and \texttt{RV64}~\cite{XKCPXKCP2026}.
Memory requirements were largest for HQC-5, ranging from 6 KB to 36\,KB ("ROM", \texttt{text} and \texttt{rodata} sections) and from 30\,KB to 45\,KB ("RAM", \texttt{sbss}, \texttt{bss}, \texttt{sdata}, \texttt{data} sections, and maximum stack).
The hardware design was described in SystemVerilog and implemented on the Xilinx Artix~7 FPGA using Vivado 2024.2.

\begin{figure}[h]
  \centering
    \includegraphics[
      width=\columnwidth,
    ]{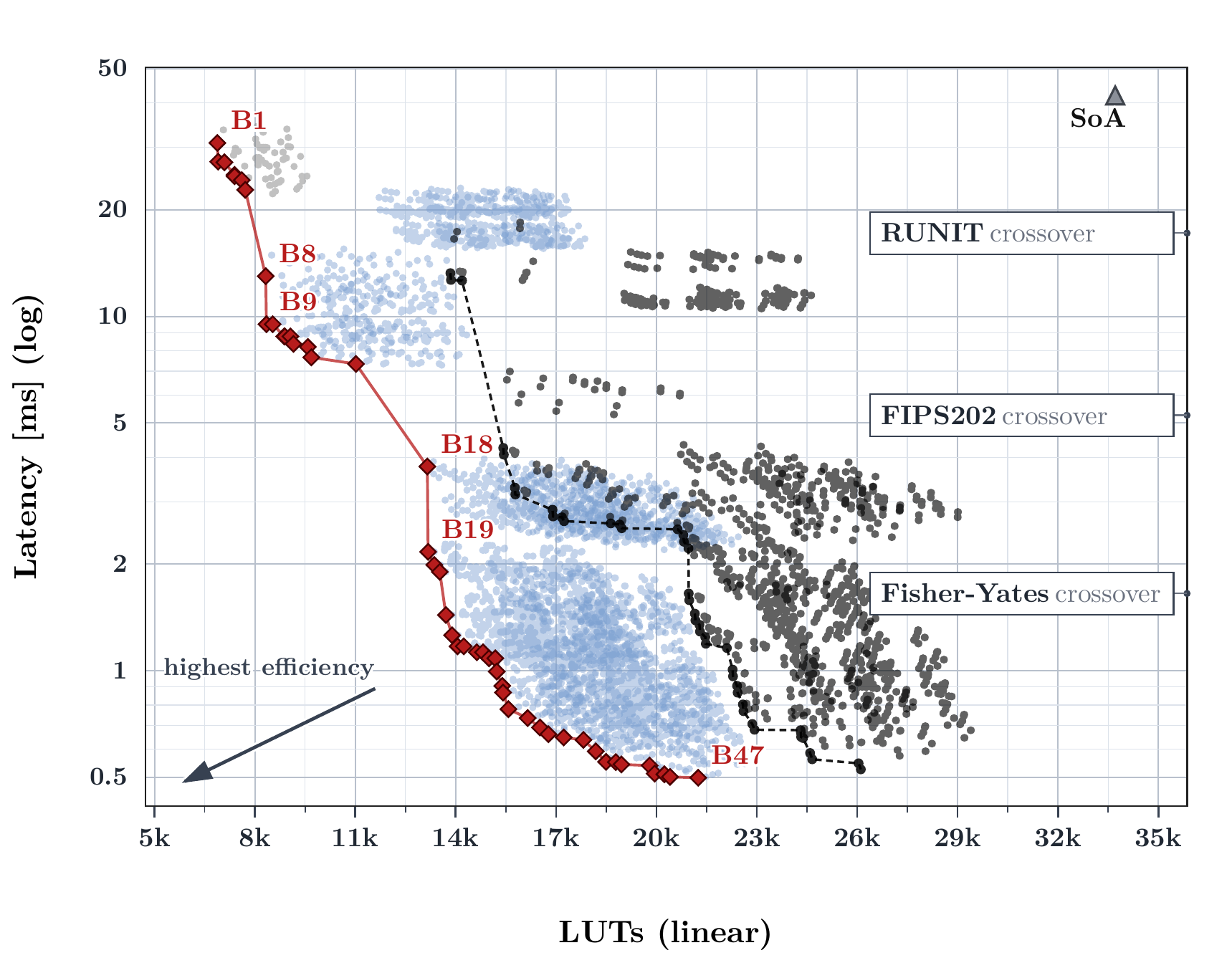}
    \caption{Bob metric (\textsc{Encaps}). Red Points - Pareto front. Blue Points - \texttt{RV32}. Black Points - \texttt{RV64}. \textbf{B1}: \texttt{RV32E}, 32b Inter. \textbf{B8}: B1 with 32b \textsc{R-Unit}, \textbf{B9}: B8 with \texttt{zbasic}, \textbf{B18}: B8 with \textsc{Fips}~202 (blocks, 16 slices), \textbf{B19}: B18 with \texttt{zbasic}, \textbf{B47}: \texttt{RV32IM\_Zpqc}, 64b inter., 128b \textsc{R-Unit}, FY, \textsc{Fips}\,202 (bytes, full state). SoA refers to~\cite{dolmetaTYRCARISCVTightlyCoupled2025}.}
    \label{fig:bob_hqc1}%
\end{figure}

As exhaustive simulation and physical implementation of all configurations is infeasible, we employ a two-stage characterization flow.
First, the \textbf{clock cycles} of the KEM functions were measured using superset systems containing all required ISA extensions and accelerators. For each measurement, the corresponding superset was loaded on the FPGA and the software was compiled for the selected ISA and accelerator configuration.
All 198,000 KEM-operation results were verified against the reference implementation on the host server.

Second, \textbf{resource utilization} was reconstructed from implementation sweeps of the RISC-V, HQC, and \textsc{Fips}~202 groups in Table~\ref{tab:configuration-space}. 
All 80 RISC-V configurations were implemented, while the HQC and \textsc{Fips}~202 sweeps used a fixed minimum \texttt{RV32E} processor and were conducted for 32- and 64-bit interconnects. 
Resource contributions of individual configurations were derived from the differences within these sweeps and combined to estimate full-system utilization. 
This was cross-verified against 200 full-system implementations distributed across the design space, yielding a mean LUT discrepancy of 2.56\,\% and a maximum discrepancy of 5.42\,\%.
Timing was evaluated over the sweeps; the critical path remained in the processor, yielding achievable frequencies ranging from 76\,MHz for the largest \texttt{RV64} configuration to 102\,MHz for the minimum \texttt{RV32E}.

LUTs were used as the primary metric for the FPGA resources, and the Flip-Flop usage stayed below 12.5K (4.7\,\% utilization).
Required Block Random Access Memories (BRAMs) are detailed in Section V-C.

\begin{figure*}[!h]
    \centering
    \includegraphics[
        width=\linewidth
    ]{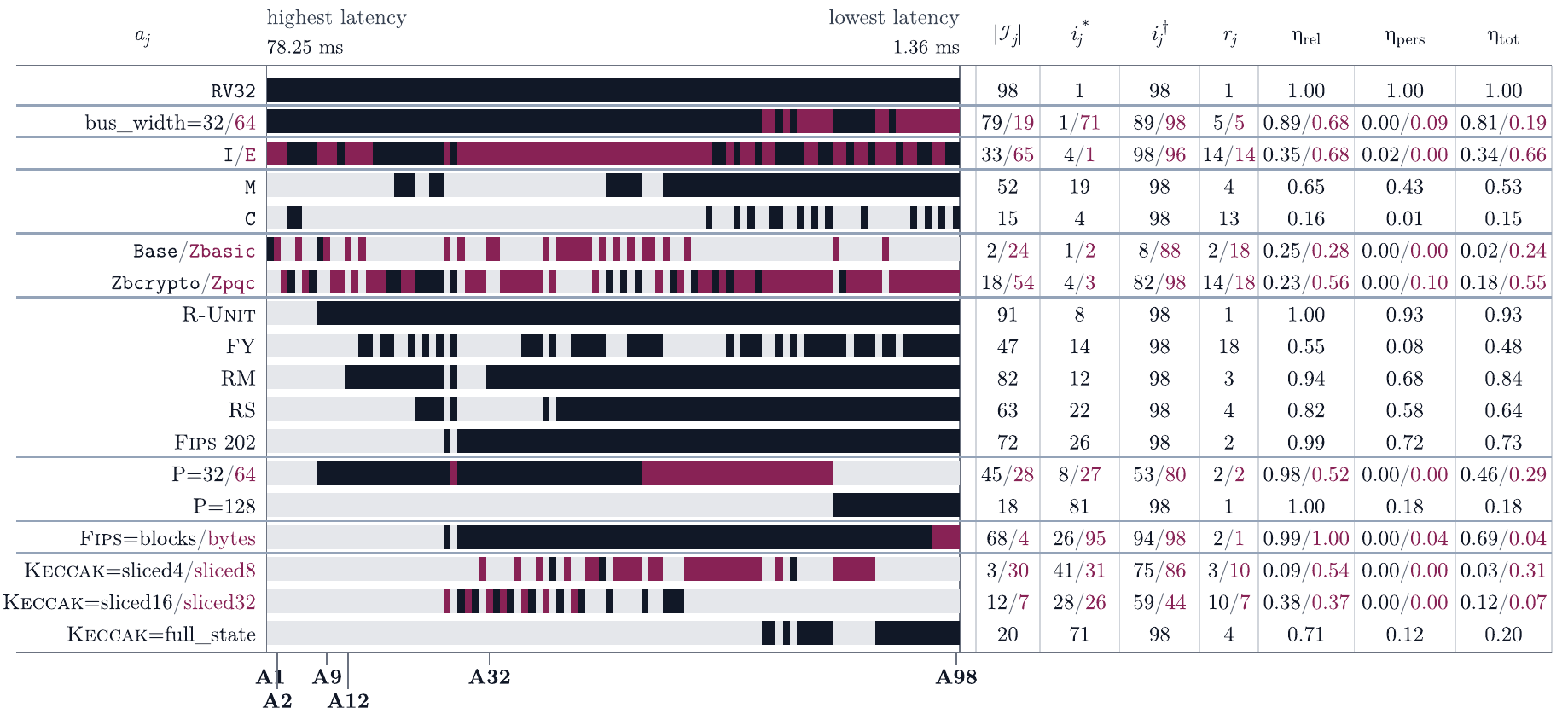}
    \caption{Occupancy statistics computed using the described evaluation methodology. For each Pareto-optimal design point in Fig.~\ref{fig:alice_hqc-1}, the figure indicates which architectural options are active; each column represents one configuration. Selected mutually exclusive options are combined into a single row and distinguished using black and wine red. The columns on the right report the corresponding occupancy metrics for each architectural option.}
    \label{fig:occupancy}
\end{figure*}

\begin{figure}
  \centering
    \includegraphics[
      width=\columnwidth,
    ]{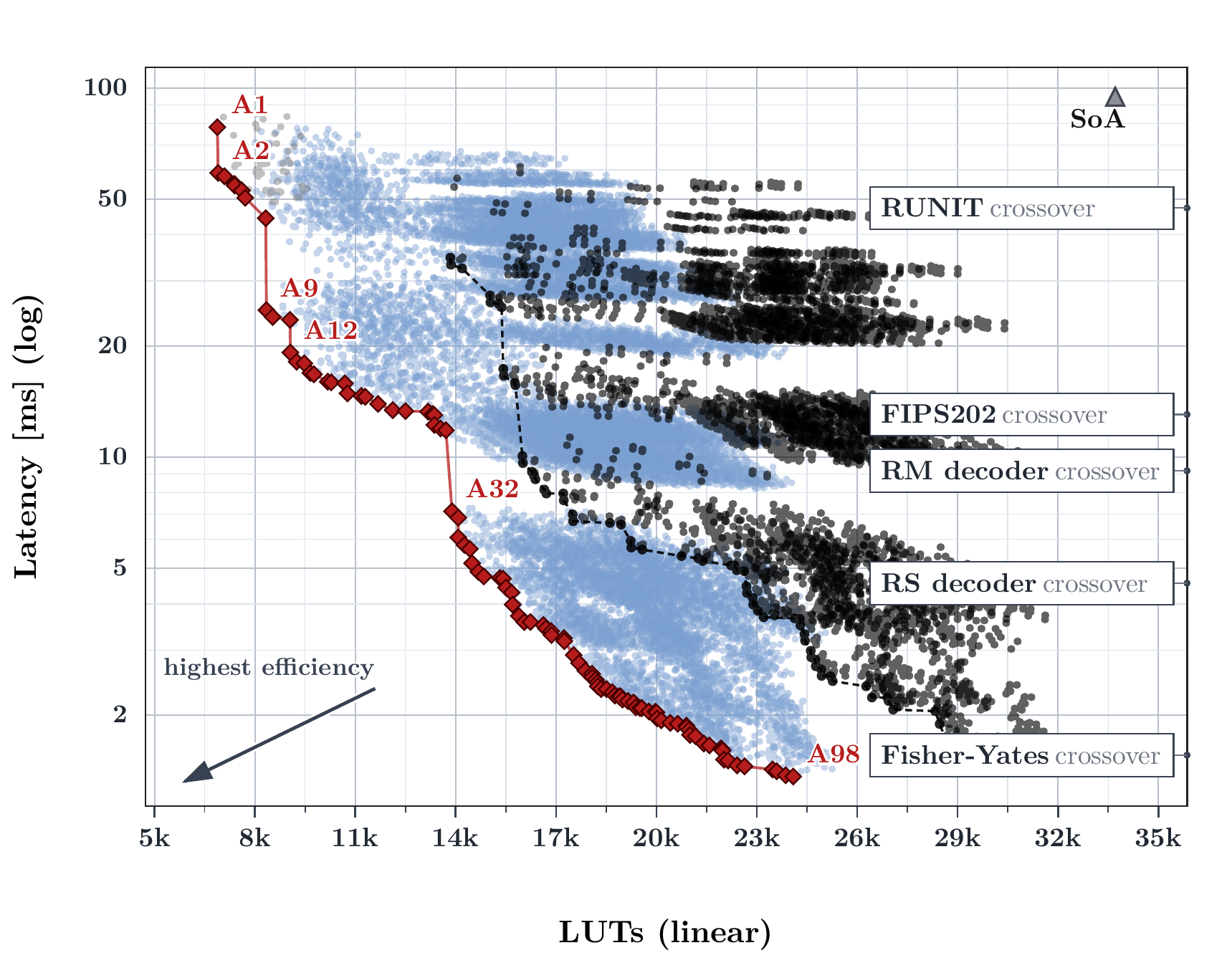}
  \caption{Alice metric (\textsc{KeyGen}~+~\textsc{Decaps}); configurations are detailed in Fig.~\ref{fig:occupancy}. Red Points - Pareto front. Blue - \texttt{RV32}. Black - \texttt{RV64}. SoA refers to~\cite{dolmetaTYRCARISCVTightlyCoupled2025}}
    \label{fig:alice_hqc-1}%
\end{figure}

\subsection{Pareto Analysis}
Figs.~\ref{fig:bob_hqc1} and~\ref{fig:alice_hqc-1} show the LUT--latency results of the explored HQC-1 configurations for the Bob and Alice metrics. Each point represents one architectural configuration. \texttt{RV32} configurations are shown in blue, whereas \texttt{RV64} configurations are shown in black. The configurations on the global LUT--latency Pareto front are highlighted in red.

As shown, the Pareto fronts range from 6.8\,kLUTs to 24\,kLUTs and reduce the latency by 57.5$\times$ from 78.25\,ms to 1.36\,ms (Alice), and 7--20.4\,kLUT at 30.86--0.50\,ms (Bob).
Overall, the explored configurations enable a fine-grained trade-off between LUTs and latency through the densely populated front; the median increment between adjacent nondominated points is $\approx$\,160 LUTs for the Alice metric.

The Pareto fronts consist of regions with moderate slopes, separated by sharp latency reductions caused by accelerator integration. The crossover markers indicate the beginning of the corresponding persistence range $\mathcal{Q}_j$: below each marker, no Pareto-optimal configuration exists without the respective accelerator. 
They therefore define an integration order, with the \textsc{R-Unit} becoming the first mandatory accelerator for both metrics.

Fig.~\ref{fig:occupancy} details the architectural choices along the Alice Pareto front. All explored features contribute to Pareto optimality except \texttt{RV64}, which is excluded because of its resource overhead and lower frequency. The 32-bit interconnect is selected in most configurations ($\eta_{\mathrm{tot}}=0.81$), whereas the base-ISA choice varies considerably, with \texttt{RV32E} appearing in approximately two-thirds of the points. The \texttt{M} extension becomes increasingly relevant toward lower latencies. The \texttt{Zpqc} profile is selected in $\eta_{\mathrm{tot}}=0.55$ of the Pareto-optimal configurations, demonstrating the benefit of its fine-grained $GF(2^8)$ instructions over the general-purpose carry-less multiplication provided by \texttt{Zbc}.

Among the accelerators, the \textsc{R-Unit} exhibits the longest persistence range ($\eta_{\mathrm{pers}}=0.93$), followed by the \textsc{Fips}~202 accelerator. In contrast, the FY sampler has the highest  $r_j$, hence mainly contributes to local refinements of the Pareto front. The \textsc{Keccak} variants provide additional fine-tuning.
The block-level \textsc{Fips}~202 interface dominates most of the Pareto front.
Beyond a distinct transition point, the byte-level variant becomes Pareto-optimal and remains so ($\eta_\text{rel}=1$) for all lower-latency configurations.

\subsection{Security Trade-Off}
\begin{figure}[t]
  \centering
  \includegraphics[
    width=0.95\columnwidth
  ]{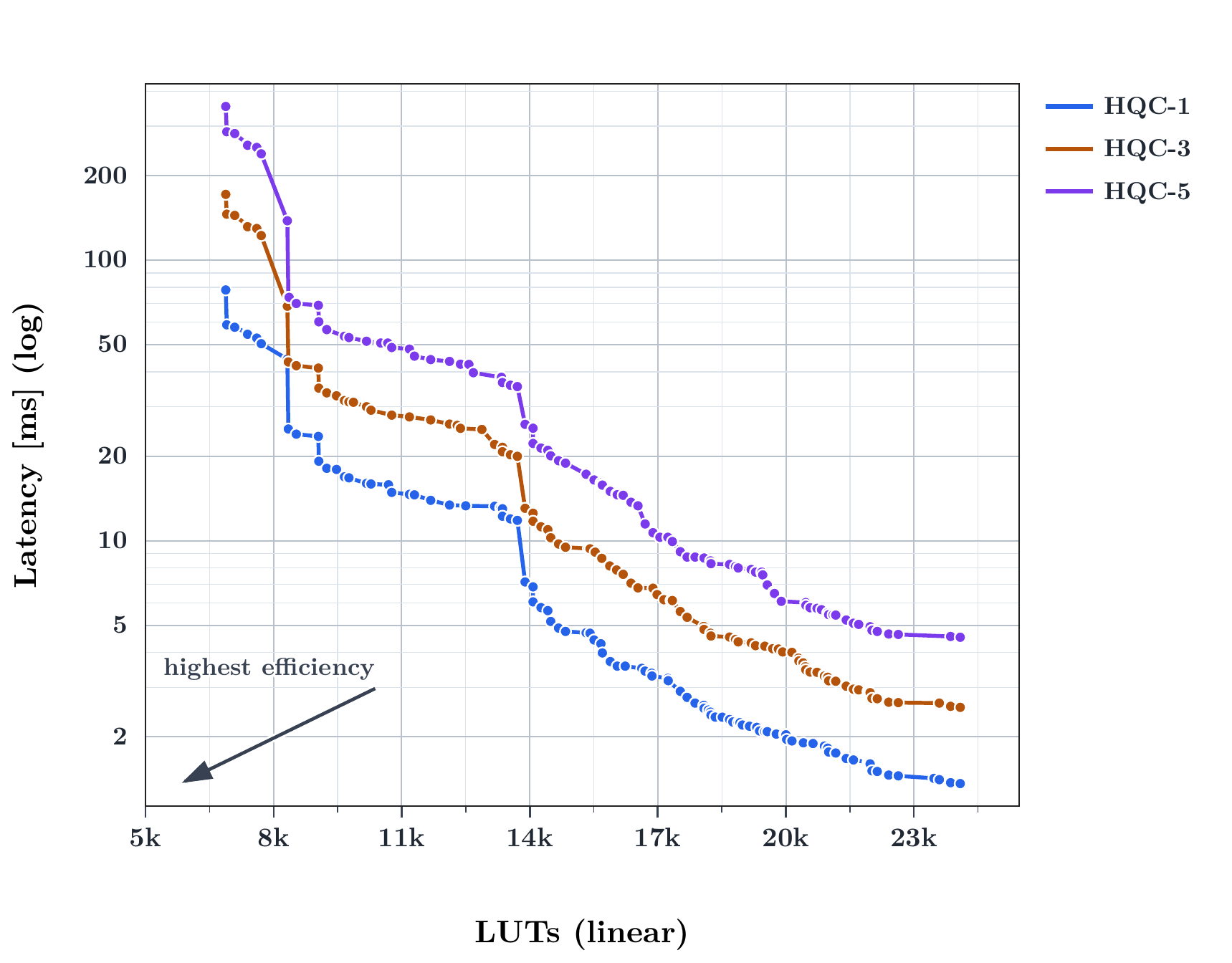}
  \caption{LUT--latency Pareto fronts for the Alice metric of the different HQC security levels.}
  \label{fig:alice_security}
\end{figure}

Fig.~\ref{fig:alice_security} shows the Pareto fronts for the different HQC security levels for the Alice metric.
Absolute latency increases with the security level, while the overall shape and accelerator ordering of the Pareto fronts are the same for different security levels. Hence, the same acceleration hierarchy applies across all parameter sets.
%, enabling application-specific trade-offs among FPGA resources, latency, and security.
For example, to guarantee an Alice latency of at most 20\,ms across all  three security levels, the \textsc{R-Unit}, RM decoder, \texttt{Zpqc} profile, and \textsc{Fips}~202 accelerator are required.
Overall, the explored architecture therefore enables application-specific selection across three dimensions: FPGA resources, latency, and required security level.

\subsection{Memory}
\begin{table}[t]
\centering
\caption{HQC-5 code and data memory for selected architectural configurations. Each entry reports the memory required for the RV32I baseline with the indicated feature enabled individually. \# denotes the number of BRAM36 equivalents.}
\label{tab:hqc5_memory}
\begin{tabular}{lccc}
\hline
\textbf{Configuration} &
\textbf{ROM [kB]} &
\textbf{RAM [kB]} &
\textbf{Total \#} \\
\hline
RV32I                 & 35.0  & 44.0 & 26 \\
+ RV32E               & 36.4 & 44.0 & 26 \\
\textbf{+ \texttt{C} }         & 29.2  & 44.0 & \textbf{24} \\
\textbf{+ FIPS~202}            & 22.0  & 44.0 & \textbf{22} \\
\textbf{+ RS decoder }         & 22.5  & 44.0 & \textbf{22} \\
+ \textsc{R-Unit}     & 35.3  & 31.2  & 22 \\
\hline
\textbf{All active} & 5.9   & 30.5  & \textbf{10} \\
\hline
\multicolumn{4}{l}{Add 2\,\# ($P=32$)/4\,\# ($P=\{64,128\}$) for HQC Accelerators.}
\end{tabular}
\end{table}
Table~\ref{tab:hqc5_memory} shows how the explored configurations affect the code (\textsc{Mem0} in Fig.~\ref{fig:architecture}) and data (\textsc{Mem1} in Fig.~\ref{fig:architecture}) memory requirements for HQC-5.
For code memory, several configurations provide substantial reductions. In particular, the \texttt{C} extension reduces the ROM requirement by approximately 6\,kB, while the \textsc{Fips}~202 accelerator reduces it by approximately 13\,kB.
Overall, processor code/data memory is reduced from 26 to 10 BRAM36 blocks, i.e., by more than 50\,\%; however, at the cost of 2 ($P=32$) or 4 ($P=64/128)$ additional BRAM36 blocks for the HQC accelerators.

\subsection{Comparison with SoA}
Table~\ref{tab:hqc_soa_comparison} shows a comparison with SoA FPGA implementations of HQC.
Design~\cite{schoeffel2022hqc} corresponds approximately to one point on our Pareto front, but implements an older HQC version.
Compared to the tightly coupled RISC-V implementation~\cite{dolmetaTYRCARISCVTightlyCoupled2025}, all of our configurations require fewer LUTs and achieve a lower latency, ranging from 4.8$\times$ lower LUT utilization at 17\,\% lower latency to 69.6$\times$ lower latency at 30\,\% lower LUT utilization (Alice metric).
A major reason is likely our use of support-form representations and our combination of custom instructions with dedicated accelerators, rather than the custom Karatsuba-based multiplication instruction used in~\cite{dolmetaTYRCARISCVTightlyCoupled2025}.
The full-hardware design~\cite{deshpandeFastEfficientHardware2024} lies below the RISC-V front in the low latency region; however, it uses an older HQC version, supports only one parameter set, and does not provide software programmability.
With respect to memory resources, our implementations require substantially fewer BRAMs than all compared prior works despite supporting all HQC parameter sets, with the most memory-efficient configuration fitting into 14 BRAMs compared to 148 BRAMs in~\cite{dolmetaTYRCARISCVTightlyCoupled2025}, corresponding to a 10.6$\times$ reduction.

\begin{table}[t]
\centering
\caption{Comparison with related HQC FPGA implementations.
Resources are reported as LUTs/FFs/\#, with LUTs and FFs given in thousands, and \# denoting BRAM36.}
\label{tab:hqc_soa_comparison}

\begin{tabular}{llrcrr}
\hline
\textbf{Work} &
\textbf{HQC} &
\textbf{Resources} &
\textbf{Key} & 
\textbf{Enc} &
\textbf{Dec} \\
&
&
\textbf{[kLUT/kFF/\#]} &
\textbf{[ms]} &
\textbf{[ms]} &
\textbf{[ms]} \\
\hline

\textbf{Ours A1}
    & 1 & 7/4/26 &
    14.43 & 30.86 & 63.82 \\

HW/SW~\cite{dolmetaTYRCARISCVTightlyCoupled2025}$^{\dagger}$
    & 1 & 34/22/148 &
    20.73 & 42.02 & 73.97 \\

HW/SW~\cite{seoHardwareSoftwareCoOptimization2026}\textsuperscript{§}
    & 1 & 17.3/8.6/32 &
    11.44 & 21.54 & 37.47 \\

\textbf{Ours A98}
    & 1 & 24.1/11/14 &
    0.25 & 0.49 & 1.11 \\

\midrule 

\textbf{Ours A1}
    & 5 & 7/3/26 &
   67.90 & 154.79 & 285.04 \\

\textbf{Ours A98}
    & 5 & 24.1/11/14 &
    0.89 & 1.78 & 3.64 \\

HW/SW~\cite{dolmetaTYRCARISCVTightlyCoupled2025}$^{\dagger}$
    & 5 & 34/22/148 &
    134.36 & 276.46 & 433.33 \\

\midrule
\multicolumn{6}{c}{\textbf{Older Version}} \\

HW/SW$^{\ddagger}$~\cite{schoeffel2022hqc}
    & 1 & 14/7/27 &
    0.56 & 1.31 & 5.57 \\

Full HW$^{*}$~\cite{deshpandeFastEfficientHardware2024}
    & 1 & 15.3/7.3/24 & 
    0.1 & 0.2 & 0.29 \\
\hline
\multicolumn{6}{l}{\footnotesize $^{\dagger}$: more powerful FPGA (Kintex-7). } \\
\multicolumn{6}{l}{\footnotesize $^{*}$: Full-hardware implementation, older version (only HQC-1 counterpart).} \\
\multicolumn{6}{l}{\footnotesize §: Assumed 100 MHz clock; true clock frequency not reported.}
\end{tabular}

\end{table}

\section{Conclusion}
This work presented, to the best of our knowledge, the most comprehensive cross-layer design-space exploration of RISC-V-based HQC implementations on FPGA, yielding both the smallest and fastest RISC-V-based HQC implementations reported to date in terms of FPGA resources and latency.
We proposed an architecture template that produces densely populated LUT--latency Pareto fronts on the Xilinx Artix 7 FPGA, enabling fine-grained architecture selection under application-specific constraints.
Using our evaluation methodology, we identified how individual features of this template shape the Pareto front and distinguished features with a persistent impact across large regions from those that provide only local fine-tuning.
All investigated architectural features appeared in at least one Pareto-optimal configuration, except for the 64-bit RISC-V variant.
Future work should extend the explored processor space, e.g., by including the vector extension, and investigate multi-algorithm PQC platforms.

\clearpage
\newpage
\bibliographystyle{IEEEtran}
\bibliography{references}

\end{document}